\documentclass[aps,prl,twocolumn,preprintnumbers]{revtex4-2}
\usepackage{amsmath, mathrsfs, amssymb,amsfonts,amsthm,graphicx, epsf, dcolumn, yfonts, subfigure}

\usepackage{etoolbox}
\apptocmd{\sloppy}{\hbadness 10000\relax}{}{}

\usepackage{kantlipsum}
\usepackage{comment}
\usepackage{graphicx,amssymb,amsmath,amsthm,amsfonts,epsfig,epsf,array,subfigure,appendix,booktabs}
\usepackage[usenames]{color}
\usepackage[unicode=true,pdfusetitle, bookmarks=true,bookmarksnumbered=false,bookmarksopen=false, breaklinks=true,
colorlinks=true,citecolor=blue]
{hyperref}
\usepackage{xr}
\usepackage[normalem]{ulem}

\def\be{\begin{equation}}
\def\ee{\end{equation}}
\def\bea{\begin{eqnarray}}
\def\eea{\end{eqnarray}}
\def\bfla{\begin{flalign}}
\def\efla{\end{flalign}}

\def\gsim{\, \rlap{$>$}{\lower 1.1ex\hbox{$\sim$}}\,}
\def\lsim{\, \rlap{$<$}{\lower 1.1ex\hbox{$\sim$}}\,}

\makeatletter
\renewcommand{\@fnsymbol}[1]{\ifcase#1\or a\or b\or c\or d\or e\or f\or g\or h\or i\or j\or k\else\@ctrerr\fi}
\makeatother

\definecolor{purple}{rgb}{0.7,0,1}

\definecolor{green}{rgb}{0,0.7,0.2}

\newcommand{\rH}{r_\text{H}}

\newcommand{\ri}{r_\text{inj}}

\newcommand{\ta}{\tilde a}

\newcommand{\dd}{\mathop{}\!\mathrm{d}}

\begin{document}

\title{Weak Cosmic Censorship with spinning particles in Kerr-(A)dS spacetimes}
\author{Antonia M.~Frassino}
\email{afrassin@sissa.it}
\affiliation{SISSA, International School for Advanced Studies, via Bonomea 265, 34136 Trieste, Italy}
\affiliation{INFN, Sezione di Trieste, via Valerio 2, 34127 Trieste, Italy}

\author{Jorge V.~Rocha}
\email{jorge.miguel.rocha@iscte-iul.pt}
\affiliation{Departamento de Matem\'atica, ISCTE--Instituto Universit\'ario de Lisboa, Avenida das For\c{c}as Armadas, 1649-026 Lisboa, Portugal}
\affiliation{Centro de Astrof\'isica e Gravita\c{c}\~ao--CENTRA, Instituto Superior T\'ecnico--IST, Universidade de Lisboa--UL, Av.\ Rovisco Pais 1, 1049-001 Lisboa, Portugal}
\affiliation{Instituto de Telecomunica\c{c}\~oes–-IUL, Avenida das For\c{c}as Armadas, 1649-026 Lisboa, Portugal}

\author{Andrea P. Sanna}
\email{asanna@roma1.infn.it}
\affiliation{INFN
Sezione di Roma, Piazzale Aldo Moro 5, 00185, Roma, Italy}
\date{\today}
\begin{abstract}
We investigate the weak cosmic censorship conjecture by analyzing the dynamics of spinning timelike particles dropped along the rotational axis of an extremal Kerr–(anti-)de Sitter black hole. 
This idea was first considered in a seminal paper by Wald and later by Needham but both analyses were restricted to asymptotically flat spacetimes.
We generalize these studies, involving spinning particles, to rotating spacetimes with non vanishing cosmological constant.
We examine whether the absorption of such particles can overspin the black hole beyond extremality, potentially leading to the formation of a naked singularity.
In asymptotically de Sitter spacetime, we find that particles that are captured cannot overspin the black hole. 
Similar conclusions hold also with anti-de Sitter asymptotics, but the analysis is more subtle, requiring careful consideration of the point particle approximation.
\end{abstract}

\maketitle

\renewcommand*{\thefootnote}{\arabic{footnote}}

\noindent \textbf{Introduction.}
The possible appearance of spacetime singularities in General Relativity, regions where Einstein's field equations break down, poses an important problem in the deterministic evolution of the spacetime geometry~\cite{Penrose:1964wq}. 
To preserve the theory’s predictive power, it is crucial that such singularities do not influence regions accessible to distant observers. This is the core idea behind the weak cosmic censorship conjecture (WCCC), formulated by Penrose~\cite{Penrose:1969pc}. It posits that, under generic initial conditions and for physically reasonable matter, singularities arising from gravitational collapse are always hidden from view to distant observers by an event horizon. 

A definitive proof of WCCC remains elusive, prompting extensive scrutiny in a variety of physical scenarios. One particularly compelling approach involves assessing whether the event horizon of a black hole (BH) can be destroyed by injecting test particles, a method pioneered by Wald~\cite{Wald:1974hkz} for an extremal Kerr-Newman BH.  
This work was later generalized by Sorce and Wald~\cite{Sorce:2017dst}, who developed a formalism that includes arbitrary matter distributions and accounts for backreaction effects at quadratic order. Their analysis confirmed that the WCCC holds for both extremal and nearly extremal Kerr-Newman BHs and provided a more robust framework for probing the conjecture.

Realistic BHs, however, are not isolated objects in asymptotically flat spacetime. Observations indicate that our universe is undergoing accelerated expansion, consistent with the presence of a positive cosmological constant and a de Sitter (dS) background~\cite{SupernovaSearchTeam:1998fmf,SupernovaCosmologyProject:1998vns,Planck:2018vyg}. Meanwhile, asymptotically anti-de Sitter (AdS) spacetimes ---characterized by a negative cosmological constant--- have become central to theoretical physics due to their role in the gauge/gravity duality~\cite{Maldacena:1997re,Aharony:1999ti}. Thus, understanding the fate of cosmic censorship in rotating BHs within both dS and AdS backgrounds is of paramount importance. Various studies have examined the conjecture in AdS settings under diverse conditions~\cite{Rocha:2011wp,Zhang:2013tba,Rocha:2014gza,Rocha:2014jma,Gwak:2015fsa,Natario:2016bay,Song:2017mdx,Gwak:2018akg,Zeng:2019aao,Chen:2019pdj,Frassino:2024fin}, as well as in the Kerr-dS case~\cite{Zhang:2013tba,Gwak:2018tmy,Natario:2019iex,Wu:2024ucf}, generally finding that the WCCC remains intact \footnote{Nevertheless, exceptions have been reported: in particular, charged models in AdS$_4$ have suggested that regions with arbitrarily large curvatures become visible to boundary observers in certain fine-tuned configurations~\cite{Crisford:2017zpi,Horowitz:2016ezu}. Additionally, small Kerr-AdS suffer from the superradiant instability, which also might impact the horizon stability and, thus, the WCCC~\cite{Cardoso:2004hs,Cardoso:2006wa,Uchikata:2009zz,Niehoff:2015oga}.}. 

\begin{figure}[t]
\centering
\vspace{-0.5cm}
\includegraphics[width=0.25\textwidth]{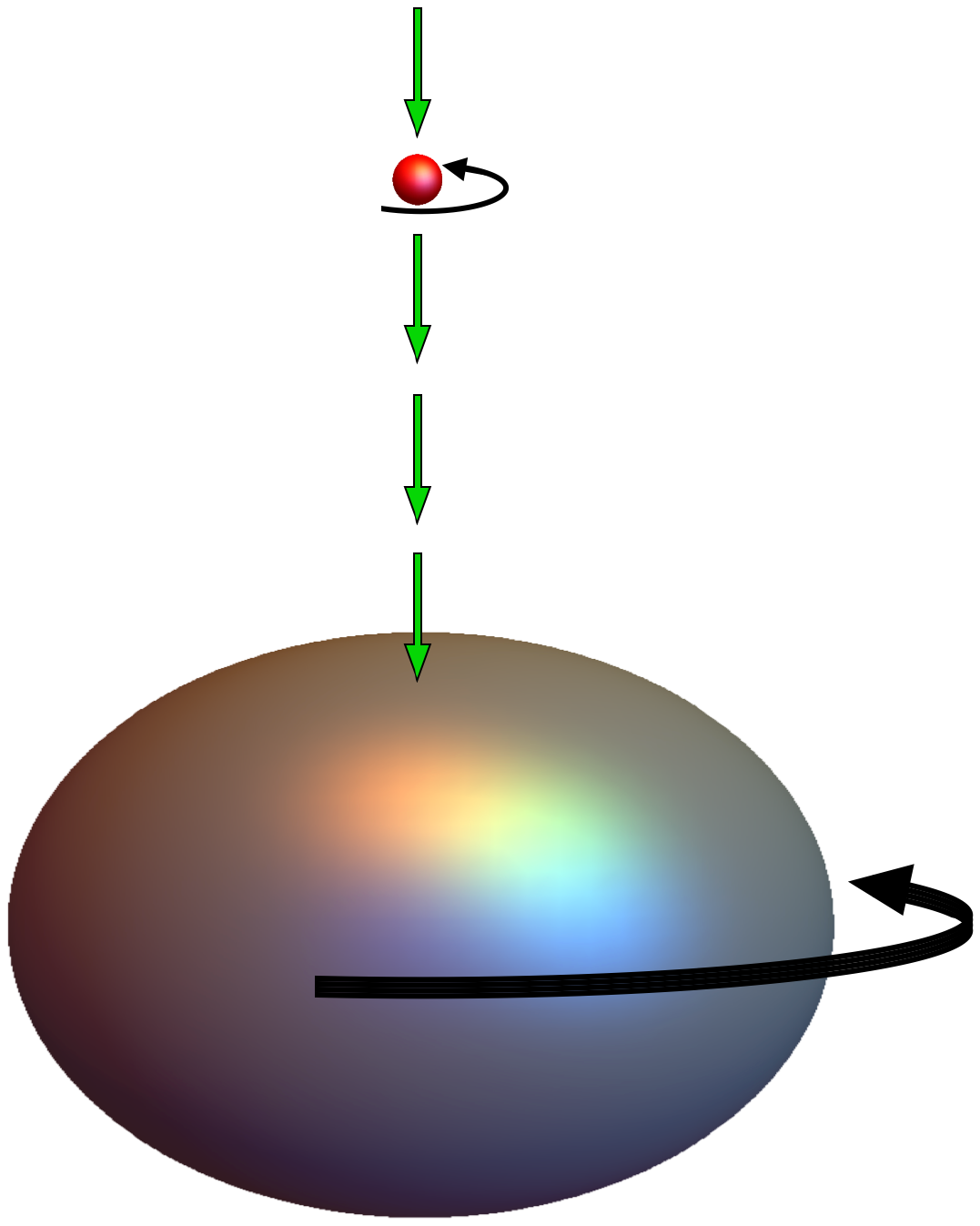}
\vspace{-0.4cm}
\caption{Schematic representation of the setup considered. A point-like spinning particle is injected and moves radially along the axis of rotation of an extremal Kerr-(A)dS BH.
}
\label{fig:Setup}
\end{figure}

In this letter, we generalize Wald's  original thought experiment to assess whether spinning particles, dropped into an extremal Kerr-(A)dS BH along its axis of rotation, 
can breach the extremality condition, effectively destroying the horizon and violating the WCCC. See Fig.~\ref{fig:Setup} for a pictorial representation of the setup considered. 
Refs.~\cite{Natario:2016bay, Natario:2019iex} proved more generally that extremal Kerr-Newman-(A)dS cannot be overcharged/overspun with test fields obeying the null energy condition. Since physical spinning bodies ---of which spinning particles can be considered a limiting case--- should satisfy the dominant energy condition \cite{Costa:2014nta}, and that implies the null energy condition, the analysis we present falls under the conditions of the theorems proved in~\cite{Natario:2016bay, Natario:2019iex}. Nevertheless, the specific calculation we perform is useful to compare with similar tests in rotating BH backgrounds different from Kerr-(A)dS.

Despite the extensive exploration of the WCCC via test particles in geodesic motion, studies involving spinning particles are scarce, and have only been carried out in asymptotically flat spacetimes~\cite{Wald:1974hkz,Tod:1976ud,Needham:1980fb}, to the best of our knowledge. Additionally, the Sorce-Wald method has not yet been applied to rotating and non-asymptotically flat BHs. Only a handful of results exist in static charged models with a cosmological constant~\cite{Yoshida:2024txh,Zhang:2020txy,Wang:2019bml}.
This gap in the literature highlights the need for further investigation in cosmologically/astrophysically relevant settings and with important holographic implications.

\noindent \textbf{Kerr-(A)dS metric.}
We consider the Kerr-(A)dS metric written in Boyer-Lindquist coordinates $(t, \, r, \, \theta, \, \varphi)$~\cite{Carter:1968ks,Carter:1973rla} (we adopt units in which $c = G = 1$)
\begin{eqnarray}
    \dd s^2 &=& -\frac{\Delta_r}{\rho^2}\left(\dd t-\frac{a \sin^2 \theta}{\Xi}\dd\varphi \right)^2 + \frac{\rho^2}{\Delta_r}\dd r^2 + \frac{\rho^2}{\Delta_\theta}\dd \theta^2 \nonumber\\
    && + \frac{\sin^2\theta\, \Delta_\theta}{\rho^2}\left(a \dd t -\frac{r^2+a^2}{\Xi}\dd \varphi \right)^2\, ,
\label{KerrAdS4D}
\end{eqnarray}
with the metric functions defined as 
\begin{eqnarray}
    \Delta_r &=& (r^2+a^2)\left(1\pm\frac{r^2}{L^2} \right) -2 M r\, ,\\
    \Delta_\theta &=& 1\mp\frac{a^2}{L^2}\cos^2\theta\, ,\\
    \rho^2 &=& r^2 + a^2 \cos^2\theta \, , \quad \Xi = 1\mp\frac{a^2}{L^2}\, ,
\label{KerrAdSMetricFunctions}
\end{eqnarray}
where $L$ is the (A)dS length related to the cosmological constant, while $M$ and $a$ denote the BH's mass and spin (per unit mass) parameters, respectively. The factor $\Xi$ ensures the correct normalization of the angular coordinate $\varphi$ to $2\pi$, thereby eliminating conical deficits at the poles $\theta = 0, \, \pi$. 

These spacetimes are solutions to the Einstein equations with a cosmological constant.
Throughout this work, the upper (lower) sign refers to the Kerr-AdS (Kerr-dS) case. The two metrics are related via the analytic continuation $L^2 \to -L^2$. The Kerr spacetime is recovered in the limit $L^2 \to \infty$.

This family of geometries is characterized by two charges, their mass and angular momentum, given respectively by~\cite{Caldarelli:1999xj,Gibbons:2004ai,Dolan:2013ft}
\begin{equation}
    {\cal M} = \frac{M}{\left(1\mp\frac{a^2}{L^2} \right)^2}\, , \qquad {\cal J} = \frac{a M}{\left(1\mp\frac{a^2}{L^2} \right)^2}\, .
    \label{KerrADScharges}
\end{equation}
Let us now briefly discuss the causal structure of these spacetimes.
Independently of the sign of the cosmological constant there is a curvature singularity at $r=0$, corresponding to a ring on the equatorial plane.
In the Kerr-dS case the function $\Delta_r$ can have at most three real roots, corresponding to the inner horizon, $r_\text{I}$, the event horizon, $\rH$, and the cosmological horizon $r_\text{C}$, satisfying $0 \leq r_\text{I} \leq \rH \leq r_\text{C}$. While the cosmological horizon is always present, the other two horizons may be absent, depending on the choice of background parameters $a$ and $M$. If these two horizons are absent, the curvature singularity is naked. The threshold separating shielded from naked singularities occurs when the event and inner horizons are degenerate, corresponding to an extremal configuration. Extremality is therefore defined by $\Delta_r(\rH) = \Delta_r'(\rH) = 0$. This is satisfied in the situation just described, i.e., $r_\text{I}=\rH$, or else in the Nariai limit, in which $\rH = r_\text{C}$. As the latter does not represent a BH spacetime, in this work we focus on the first possibility. The explicit extremality conditions read:
\begin{equation}
    M = \frac{a^2 L^2 + \rH^4}{L^2 \rH} \, , \qquad \rH  = \sqrt{\frac{L^2-a^2-\beta_-}{6}}  \,, 
\label{extremalconditionKerrdS}
\end{equation}
where we defined, for brevity, $\beta_\pm \equiv \sqrt{a^4+L^4 \pm 14a^2 L^2}$.
To ensure the reality of the above expressions in the dS case, we must restrict the BH spin parameter to the range $0 \leq a/L \leq \tilde{a}_{\rm max}$, with  $\tilde{a}_{\rm max} = 2-\sqrt{3} \simeq 0.268$ (which solves $\beta_- = 0$).
In the Kerr-AdS case, instead, the spacetime may exhibit two horizons (i.e., two real distinct roots of $\Delta_r$), no horizons (corresponding to a naked singularity), or an extremal horizon, again defined as before. The extremality condition, in this case, yields
\begin{equation}
    M = \frac{a^2 L^2-\rH^4}{L^2 \rH}\, , \qquad 
    \rH = \sqrt{\frac{\beta_+-a^2-L^2}{6}}\, . 
\label{extremalityconditionKADS}
\end{equation}
Furthermore, we restrict the parameters to $0 \leq a/L < 1$, in order to avoid divergences in the line element~\eqref{KerrAdS4D} \footnote{The restriction to non-negative values for $a/L$ is only a matter of convention. Negative values can be considered by flipping the sign of $a$.}.

In the limit $L^2 \to \infty$, both Eqs.~\eqref{extremalconditionKerrdS} and~\eqref{extremalityconditionKADS} reduce to $\rH = M = a$, as expected, since the latter is the extremality condition for Kerr BHs.

\noindent \textbf{Spinning test particles along the axis of rotation.}
The equations describing the motion of spinning test particles are known as the Mathisson-Papapetrou-Dixon (MPD) equations~\cite{Mathisson:1937zz,Papapetrou:1951pa,Dixon:1964cjb}:
\begin{subequations}
\begin{align}
    \frac{\text{D} p^\mu}{\text{D}\lambda} &= -\frac{1}{2} {R^\mu}_{\nu \rho \sigma} v^\nu S^{\rho \sigma}\,, \label{MPD1}\\
    \frac{\text{D}S^{\mu\nu}}{\text{D}\lambda} &= p^\mu v^\nu - p^\nu v^\mu\, ,\label{spintensoreom}
\end{align}  
\label{MPDequations}
\end{subequations}
where ${R^\mu}_{\nu \rho \sigma}$ is the Riemann tensor, $\lambda$ is an affine parameter along the trajectory, $\text{D}/\text{D}\lambda \equiv v^\nu \nabla_\nu$ is the directional covariant derivative, $v^\mu$ represents the 4-vector tangent to the trajectory while $p^\mu$ is the total 4-momentum of the particle (of which we provide more details below), and $S^{\mu\nu}$ is the antisymmetric spin tensor representing the internal angular momentum of the particle.

The dynamics described by Eqs.~\eqref{MPDequations} is not closed. A supplementary spin condition is typically imposed, fixing the center of the particle, thus specifying its world line. 
We adopt one of the most widely used choices, the Tulczyjew condition $p_\mu S^{\mu\nu} = 0$~\cite{tulczyjew1959equations}. It ensures that the intrinsic particle spin $s$, defined by
\begin{equation} \label{eq:normalization_cond}
s^2 = \frac{1}{2}S^{\mu\nu}S_{\mu\nu}\,,
\end{equation}
is conserved, by virtue of Eq.~\eqref{spintensoreom} (see also Ref.~\cite{Wald:1972sz}). 
The right hand side of Eq.~\eqref{MPD1} describes the spin-curvature interaction~\cite{Wald:1972sz,Carmeli:1976mq}.
The test particle's 4-momentum $p^\mu$ that allows to express the MPD equations in the form~\eqref{MPDequations} is defined by
\begin{equation}\label{momentumspinning}
    p^\mu = m v^\mu \, - \, v_\nu \frac{\text{D}S^{\mu\nu}}{\text{D}\lambda}\, ,
\end{equation}
where $m$ is the mass of the particle. 
It can be checked that $m^2 = -p_\mu p^\mu$ is also a conserved quantity~\cite{Wald:1972sz,Carmeli:1976mq}. 

In this letter, we perform WCCC tests considering the absorption of spinning massive test particles falling along the BH's rotation axis $\theta = 0$, with the particle spin parallel to the axis (see Fig.~\ref{fig:Setup}). In this specific setup, the tangent vector has components $v^\mu = (v^t,\, v^r,\, 0,\, 0)$, while the only non-zero components of the spin tensor are $S_{\theta\varphi} = -S_{\varphi\theta}$ (see also Ref.~\cite{Mino:1995fm}). 
From Eq.~\eqref{momentumspinning}, the 4-momentum $p^\mu$ is parallel to $v^\mu$ in this case, i.e., $p^\mu = m v^\mu = (p^t, \, p^r, \, 0, \, 0)$.
To evaluate $S^{\theta\varphi}$, we employ the normalization condition \eqref{eq:normalization_cond}, which gives
\begin{equation}
     S^{\theta \varphi} = \frac{s}{\sqrt{g_{\theta\theta}g_{\varphi \varphi}}} = \frac{1}{\sin\theta} \, \frac{s}{a^2 + r^2} \, \left(1\mp\frac{a^2}{L^2} \right) + \mathcal{O}(\theta)\, ,
    \label{SthetaphiCmetric}
\end{equation}
where in the second equality we performed a series expansion around the axis of rotation $\theta = 0$. Clearly, $S^{\theta\varphi}$ diverges on the axis, but all scalar invariants remain regular there.

In the following, we will consider future-directed particles, namely those with $p_t < 0$ \footnote{Indeed, future directed motion implies $v^t = \dot t >0$. From the fact that, for spinning particles along the axis, $p^\mu \parallel v^\mu$, future directed particles also have $p^t >0$, implying $p_t = g_{tt} p^t < 0$.}, that are captured by the BH.
To compute $p_t$, we exploit time translation invariance of the spacetime and use the conserved quantities along the motion of spinning particles, whose general expression reads as~\cite{Dixon:1970zza}
\begin{equation}\label{EconservedKerr}
    Q_{\xi} = \xi^\mu p_\mu - \frac{1}{2} S^{\mu\nu}\nabla_\nu \xi_\mu = \xi^\mu p_\mu - \frac{1}{2}S^{\mu\nu} \partial_\nu \xi_\mu \, ,
\end{equation}
where the $\xi$'s are the Killing vectors $\xi^\mu_{(t)} = \delta^\mu_t$ and $\xi^\mu_{(\phi)} = \delta^\mu_\phi$, related to time-translation and rotational invariance, respectively. $Q_{\partial_\varphi}$, related to the orbital angular momentum, is zero for spinning particles moving along the BH axis of rotation.
The energy $E$ of the particle, instead, is associated with $\xi^\mu_{(t)}$ and can be written as $-E = p_t + \frac{1}{2}S^{\theta\varphi} \partial_\theta g_{t \varphi}$, or, explicitly, 
\begin{equation}
\begin{split}
    p_t &= -E + a s \left[\frac{2M r}{\left(a^2+r^2\right)^2}\mp \frac{1}{L^2}\right]\, .
    \label{ptfromconservedcharges}
\end{split}
\end{equation}
The condition $p_t < 0$, thus, translates into an upper bound on the spin-to-energy ratio $\hat{s} \equiv s/E$ for a particle to be absorbed by the BH.
In the Kerr-dS case, imposing $p_t < 0$ at the extremal horizon and using Eq.~\eqref{extremalconditionKerrdS} yields
\begin{equation}
    \hat s < \frac{L^2 \left(5 a^2+L^2-\beta_-\right)}{6 a \left(L^2+a^2\right)} \equiv \hat{s}_{1}^{\rm dS}.
    \label{spinUBKerrdS}
\end{equation}
Similarly, in the extremal Kerr-AdS case, one obtains, using Eq.~\eqref{extremalityconditionKADS},
\begin{equation}
    \hat s < \frac{L^2 \left(5 a^2-L^2+\beta_+\right)}{6 a \left(L^2-a^2\right)} \equiv \hat{s}_{1}^{\rm AdS}\, .
    \label{spinUBKerrAdS}
\end{equation}
To test the WCCC using the standard procedure, one considers infalling particles with normalized spins $\hat{s}$ saturating these bounds. Clearly this is the case that has better chances of overspinning the BH. One then analyzes the final geometry after the particle is absorbed (assuming the perturbed BH settles down to a Kerr-(A)dS spacetime, and neglecting energy and angular momentum loss due to gravitational radiation). This approach was employed by Wald~\cite{Wald:1974hkz} for an extremal Kerr BH and it is instructive to do the comparison.
Indeed, in the $L^2 \to \infty$ limit, both Eqs.~\eqref{spinUBKerrdS} and~\eqref{spinUBKerrAdS} reduce to $\hat s < 2 a$, which matches the condition found in~\cite{Wald:1974hkz}. This bound has precisely the opposite sign compared to what is necessary to overspin an extremal Kerr BH, i.e., $\hat s > 2a$. The physical conclusion is, thus, that particles with dangerously high spins are not absorbed and that the WCCC is preserved for an extremal Kerr BH absorbing spinning particles. 

\noindent 
\emph{Validity of the point particle approximation.}
Following the described procedure, however, could often lead to apparent WCCC violations~\cite{Needham:1980fb}. Unlike point particles in geodesic motion, spinning particles require additional restrictions to be considered point-like.

First, since objects cannot spin faster than the speed of light, they must have a minimum size $r_0 \gtrsim s/m$. This can be understood from the approximate relation $s \sim m r_0 v_\text{max}$, with $v_\text{max}$ the spinning velocity, constrained by $v_\text{max} < 1$~\cite{Wald:1972sz}. Moreover, their motion is correctly described by the MPD equations as long as they are treated as point-like objects. This implies that $r_0$ must be much smaller than the BH size, set, in our context, by the extremal horizon radius $\rH$. The regime of interest is, thus, $s/m \ll \rH$.\\ 
%
\noindent \textbf{Tests of the WCCC.}
We now consider an extremal Kerr-(A)dS background perturbed by the absorption of a spinning particle falling along the axis of rotation. This perturbation induces a shift in the event horizon such that $\rH \to \rH + \delta \rH$ (with $\delta \rH \ll \rH$). For the initial extremal configuration, $\rH$ coincides with the location of the minimum of $\Delta_r$, which we denote generically by $r_m$. Simultaneously, the mass and angular momentum of the BH vary accordingly, ${\cal M} \to {\cal M} + \delta {\cal M}$ and ${\cal J} \to {\cal J} + \delta {\cal J}$, with $\delta {\cal M} \ll {\cal M}$ and $\delta {\cal J} \ll {\cal J}$ (or equivalently, in terms of the BH mass $M$ and spin parameter $a$, as $M \to M + \delta M$ and $a \to a + \delta a$). Here, $\delta {\cal M}$ and $\delta {\cal J}$  correspond to the energy $E$ and the intrinsic spin $s$ of the absorbed particle, respectively. This notation will be used henceforth.
One then considers linear perturbations of the metric function induced by the absorption of the particle and evaluates the perturbed metric function at its minimum. If the resulting value is non-positive, the WCCC is preserved; otherwise, capture of the particle disrupts the BH's horizon.

\begin{figure}[t]
\centering
\includegraphics[width=0.45\textwidth]{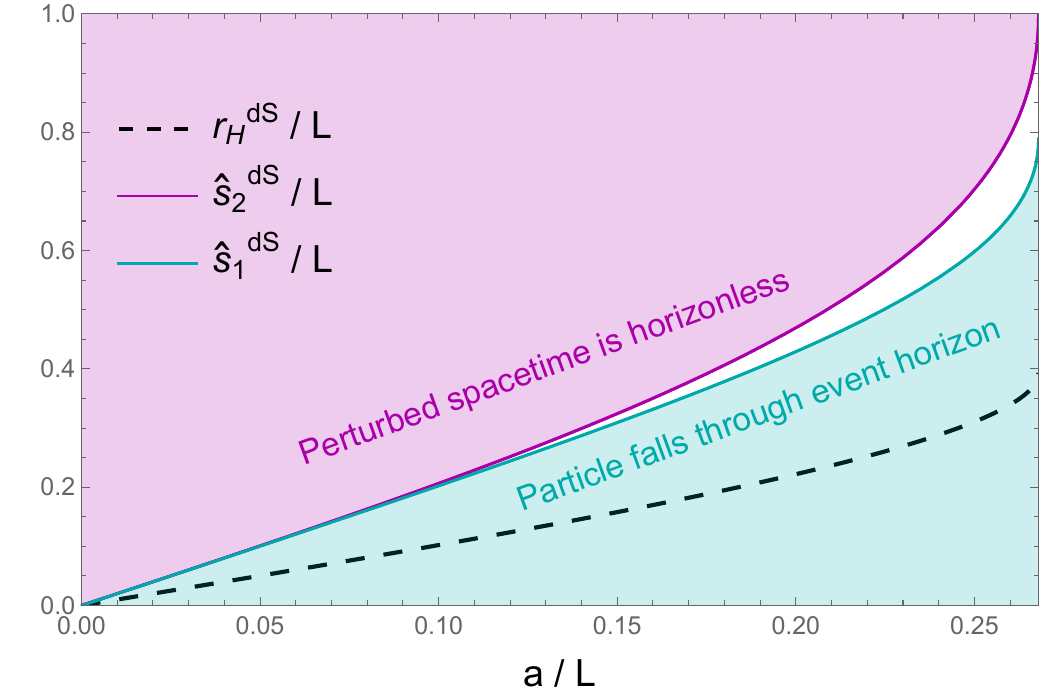}
\caption{Comparison, for the Kerr-dS case, between the upper bound on $\hat{s}/L$ (light blue solid curve) allowing the particle to be captured by the BH, given by Eq.~\eqref{spinUBKerrdS}, the lower bound on $\hat{s}/L$ (violet solid curve) required to destroy the event horizon, as indicated by Eq.~\eqref{s2KerrdS}, and the radius of the extremal BH~\eqref{extremalconditionKerrdS} (black dashed curve), which determines the regime of validity of the point-particle approximation. All three quantities are plotted as functions of $a/L$. Note that here $a/L$ is restricted to the range $[0,\tilde{a}_{\rm max}]$, in accordance with the discussion in the main text.}
    \label{fig:Comp2}
\end{figure}

In our case, the metric function, evaluated at the minimum, $r_m$, reads
\begin{equation}
    \Delta_r(r_m + \delta r, M + \delta M, a + \delta a) =  \Delta_r(r_m, M, a)
    + \delta \Delta_r \, ,
\end{equation}
where
\begin{equation}
    \delta \Delta_r =  \partial_r \Delta_r|_{r_m} \delta r + \partial_M \Delta_r|_{r_m} \delta M + \partial_a \Delta_r|_{r_m} \delta a\, . 
    \label{deltaDeltarKADSfirst}
\end{equation}
At extremality, $\Delta_r(r_\text{H}, \, M, \, a) = \partial_r\Delta_r|_{r_\text{H}} = 0$.
Expressing $\left(\delta M, \, \delta a\right)$ as functions of $\left(E, \, s\right)$ by inverting 
\begin{subequations}
\begin{align}
    E &= \frac{\partial {\cal M}}{\partial a} \delta a + \frac{\partial {\cal M}}{\partial M} \delta M\, ,\\
    s & = \frac{\partial {\cal J}}{\partial a} \delta a + \frac{\partial {\cal J}}{\partial M} \delta M\, ,
\end{align}
\end{subequations}
and using the explicit expressions for the charges~\eqref{KerrADScharges} gives (we recall our conventions: upper and lower signs refer to Kerr-AdS and Kerr-dS, respectively) 
\begin{subequations}
\begin{align}
    \delta a & = \frac{\left(L^2\mp a^2\right)^2 (s-a E)}{L^4 M}\, ,\\
    \delta M & = \frac{\left(L^2\mp a^2\right) \left[E \left(L^2 \pm 3 a^2\right)\mp 4 a s\right]}{L^4}\, .
\end{align}   
\label{deltaadeltaM}
\end{subequations}
In the Kerr-dS case, plugging Eqs.~\eqref{deltaadeltaM} into Eq.~\eqref{deltaDeltarKADSfirst}, together with Eqs.~\eqref{extremalconditionKerrdS} yields
\begin{eqnarray}
    \delta \Delta_r \!&=& \!\frac{2E \sqrt{L^2-a^2-\beta_-}\!  \left(a^2+L^2\right)}{\sqrt{6} L^4}
    \biggl[3a^2 \!-\! L^2 \!-\! 4a\hat{s}\\
    &&  -\frac{3 a L^2 (\hat{s}-a) \left(a^2+L^2\right)}{2L^2 - 2a^2 + \beta_-} \left(\frac{1}{L^2}+\frac{6}{a^2-L^2+\beta_-}\right) \biggr]\, .\nonumber
\end{eqnarray}
The pre-factor in front of the square brackets is positive, so we can focus on the enclosed expression. Requiring $\delta \Delta_r \leq 0$ (consistent with WCCC preservation) imposes another upper bound on $\hat s$ (for convenience, we also define the dimensionless parameter $\ta = a/L$)
\begin{eqnarray}
    \hat s &\leq&  
    \frac{L}{\ta} \biggl[11 \ta^4-\left( \frac{7\beta_-}{L^2}+22\right) \ta^2 +\frac{\beta_-}{L^2}-1 \biggr] \\
    && \times \biggl[\ta^4 +\left(\frac{\beta_-}{L^2}+22\right) \ta^2 -\frac{7\beta_-}{L^2}-11\biggr]^{-1} \equiv \hat s^{\text{dS}}_2\, , \nonumber
\label{s2KerrdS}
\end{eqnarray}
which must be consistent with Eq.~\eqref{spinUBKerrdS} to ensure both particle capture and horizon preservation. 
It turns out that $\hat{s}_{1}^{\rm dS} \leq \hat{s}_{2}^{\rm dS}$. This can be explicitly confirmed by considering $\frac{\tilde{a}}{L}\left(\hat{s}_{2}^{\rm dS}-\hat{s}_{1}^{\rm dS}\right)$, which has an absolute minimum equal to zero at $\tilde{a}=0$, and monotonically increases for $0<\tilde{a}<\tilde{a}_{\rm max}$. Therefore, the condition $\hat{s} < \hat{s}_{1}^{\rm dS}$, required for the particle to be captured by the BH, automatically implies the condition $\hat{s} \leq \hat{s}_{2}^{\rm dS}$ to preserve its event horizon (see also Fig.~\ref{fig:Comp2}).
Therefore, an extremal Kerr-dS BH cannot be overspun under the process considered.\\

\begin{figure}[!t]
\centering
\includegraphics[width=0.45\textwidth]{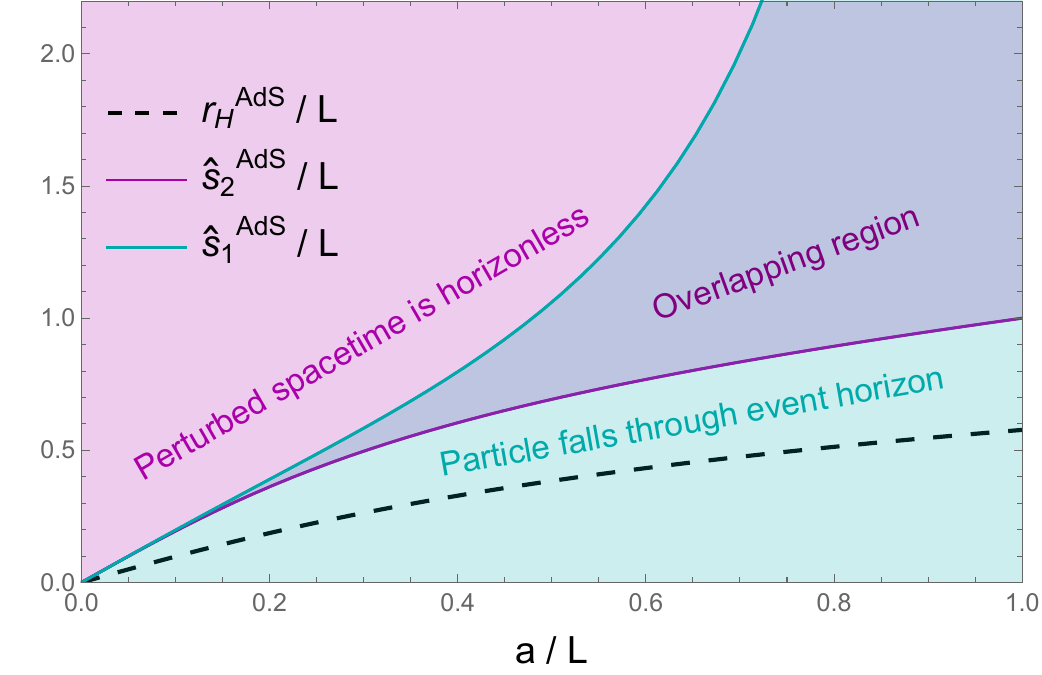}
    \caption{Comparison, for the Kerr-AdS case, between the upper bound on $\hat{s}/L$ (light blue solid curve) allowing the particle to be captured by the BH, given by Eq.~\eqref{spinUBKerrAdS}, the lower bound on $\hat{s}/L$ (violet solid curve) required to destroy the event horizon, as indicated by Eq.~\eqref{s2KerrAdS}, and the radius of the extremal BH~\eqref{extremalityconditionKADS} (black dashed curve), which determines the regime of validity of the point-particle approximation. $a/L$ is restricted to the range $[0,1)$ in accordance with the discussion in the main text. Note that in this case there is an overlap region where the particle can have sufficiently small spin to be captured by the BH but still have sufficiently large spin to destroy the event horizon.}
    \label{fig:Comp1}
\end{figure}

For the Kerr-AdS metric, instead, plugging Eqs.~\eqref{deltaadeltaM} into Eq.~\eqref{deltaDeltarKADSfirst} and using Eq.~\eqref{extremalityconditionKADS} leads to 
\begin{eqnarray}
    \delta \Delta_r &=& \frac{2 E \sqrt{\beta_+ - L^2 - a^2}(L^2 - a^2)}{\sqrt{6}L^4} \biggl[-3a^2 - L^2\\
    &&+ 4 a \hat s + 6 \frac{a(\hat s -a)(L^2-a^2)(5L^2-a^2 + \beta_+)}{36 a^2 L^2 - (L^2 + a^2-\beta_+)^2}\biggr ]\, . \nonumber
\label{DeltaDrExtremalGeneral}
\end{eqnarray}
Since the pre-factor outside the square brackets is positive, we can again focus on the expression within. Requiring $\delta \Delta_r \leq 0$ imposes again another upper bound on $\hat s$:
\begin{eqnarray}
    \hat s &\leq&  
    \frac{L}{\ta} \biggl[11\ta^4+ \left(\frac{7\beta_+}{L^2} +22\right)\ta^2 + \frac{\beta_+}{L^2}-1 \biggr] \\
    && \times \biggl[-\ta^4+\left(\frac{\beta_+}{L^2}+22 \right)\ta^2+\frac{7\beta_+}{L^2}+11\biggr]^{-1} \equiv \hat s^{\text{AdS}}_2\, . \nonumber
\label{s2KerrAdS}
\end{eqnarray}
An inspection of $\hat s^{\text{AdS}}_1 - \hat s^{\text{AdS}}_2$ reveals that this function has an absolute minimum at $\ta = 0$ and monotonically increases with $\ta$. Thus, contrary to the Kerr-dS case, in AdS one finds $\hat s^{\text{AdS}}_1 \geq \hat s^{\text{AdS}}_2$. This hints at a potential violation of the WCCC, since there is an overlap region in the parameter space where a spinning particle can be absorbed by the BH and overspin it. 
By numerically scanning the parameter space, we find that, as long as we consider particles with a sufficiently low spin-to-mass ratio to satisfy the point-particle approximation ($s/m \lsim 0.1 \, \rH$), and they are injected from a sufficiently large distance from the BH horizon ($\ri - \rH \gsim \rH$, where $\ri$ is the radius corresponding to an outer turning point at which $v^r = 0$), $\hat s/\rH$ always remains below unity. In the parameter space, these particles live below the dashed black curve in Fig.~\ref{fig:Comp1} and are thus unable to spin the BH past extremality.

A natural question then arises regarding the fate of particles injected in the immediate vicinity of the BH. In this regime, the finite size of the particle should also be compared with the distance from the horizon. Specifically, it should satisfy $r_0 < \ri - \rH$. We find that particles with fine-tuned parameters capable of disrupting the horizon possess a size exceeding this bound, implying they were already inside the BH when they were injected~\footnote{We thank J. Nat\'ario for pointing out this consideration.}. Conversely, particles smaller than $\ri - \rH$ are unable to disrupt the horizon, no matter how close from it they are injected (see the Appendix for further details).

\noindent \textbf{Conclusions and outlook.}
Our analysis extends the existing literature on testing the WCCC with spinning particles in asymptotically flat spacetimes by exploring analogous processes in the presence of a timelike boundary (AdS) and a cosmological horizon (dS). We find that an extremal Kerr–dS BH resists overspinning attempts via the absorption of spinning particles falling along its axis of rotation. Similar conclusions hold in the Kerr-AdS case, although neglecting a careful evaluation of the point particle approximation would lead to the destruction of the horizon in finely tuned scenarios. 
These results are consistent with the findings of Ref.~\cite{Natario:2016bay}. Considering different matter content and BH backgrounds, potential WCCC violations in AdS spacetimes have been suggested~\cite{Horowitz:2003yv, Horowitz:2016ezu, Crisford:2017zpi}. At any rate, our results for Kerr-AdS may offer further insights in holographic contexts~\cite{Engelhardt:2019btp,Engelhardt:2020mme,Frassino:2025buh}. 

These results also connect with broader considerations of geometric inequalities involving BH parameters. In asymptotically flat spacetimes, the WCCC is closely tied to the Penrose inequality, which relates a BH’s mass and horizon area~\cite{Penrose:1973um}. A corresponding inequality has been conjectured for AdS spacetimes~\cite{Itkin:2011ph, Frassino:2024bjg} or in dS (see, e.g., Ref.~\cite{Mars:2009cj}).
A recent proposal~\cite{Amo:2023bbo} presents a new class of inequalities for stationary asymptotically AdS BHs, relating their horizon area to other thermodynamic quantities. Our results, which highlight the importance of the point-particle approximation in the Kerr-AdS background, provide a framework to similarly test the WCCC using more general rotating BHs in AdS, including charges and/or matter fields, and offer a complementary perspective to the limits imposed by the inequalities put forward in Ref.~\cite{Amo:2023bbo}.

Finally, our analysis lays the foundation for extending these ideas to rotating and accelerated BHs, which, in combination with the braneworld scenario, admit a holographic dual description in terms of quantum-corrected BHs~\cite{Emparan:2020znc}. In this framework, the injection of a spinning particle in the bulk acquires a lower-dimensional interpretation on the brane, directly connecting with the results of Ref.~\cite{Frassino:2024fin}. 

\medskip

\noindent \emph{Acknowledgments.}
We thank J. Nat\'ario for insightful and useful discussions.
AMF acknowledges the support from EU Horizon 2020 Research and Innovation Programme under the Marie Sk\l{}odowska-Curie Grant Agreement no.~101149470.
JVR acknowledges the support from {\it Funda\c{c}\~ao para a Ci\^encia e a Tecnologia} under project 2024.04456.CERN. APS gratefully acknowledges support from a research grant funded under the INFN–ASPAL agreement as part of the Einstein Telescope training program. APS is also partially supported by the MUR FIS2 Advanced Grant ET-NOW (CUP:~B53C25001080001) and by the INFN TEONGRAV initiative.

\vspace{-0.3cm}


\bibliography{refs}

\newpage 

\begin{appendix}
\onecolumngrid

\section{Appendix}

In the following, we provide additional details concerning the calculations supporting the main text's conclusion that spinning particles moving along the axis of rotation of an extremal Kerr-AdS BH cannot overspin it, in the regime where the point particle approximation is valid.

We recall the setup considered: a particle of mass $m$ and intrinsic spin $s$ falls along the axis of rotation from an outer turning point (at which $v^r = 0$) at a radial distance $\ri$ from the event horizon $\rH$. We imagine that the particle is `injected' at that location, since the particle's low energy level combined with the confining AdS potential do not allow the particle to come in from larger radii. We will not delve into its possible origin, but such a particle could be the product of a scattering process in the vicinity of the BH, for example.

Since for a spinning particle moving along $\theta = 0$, the 4-momentum is parallel to $v^\mu$ (see the main text), we have also $p^r = 0$ at the turning point. The timelike constraint $p_\mu p^\mu = -m^2$ in this situation, thus, reduces to $g_{tt} \left(p^t\right)^2 = -m^2$. Evaluating the latter onto the axis of rotation and at $r=\ri$, considering future directed particles with $p_t < 0$ and using Eq.~\eqref{ptfromconservedcharges} in the main text gives
\begin{equation}
    \frac{E}{m} = \sqrt{\frac{\Delta_r(\ri)}{\ri^2 + a^2}} + a \frac{s}{m} \left[\frac{2M r}{(\ri^2 + a^2)^2} - \frac{1}{L^2} \right]\, .
\label{Energyovermass}
\end{equation}
Here, it is assumed that $M$ takes its extremal value (Eq.~\eqref{extremalityconditionKADS} in the main text). Since the AdS length $L$ can be scaled away and also the particle mass $m$ enters only through the ratios $E/m$ and $s/m$, we can compute the energy-to-mass ratio of the particle depending on $\ri$, $s/m$ and $a$. The spin-to-mass ratio $s/m$ is subjected to the point-particle constraint, i.e., $s/m \ll \rH$, while we restrict also to $0 \leq a/L < 1$. 

The main idea is to employ Eq. \eqref{Energyovermass} to evaluate
\begin{equation}
    \frac{\hat s}{\rH} = \frac{s}{m} \cdot \frac{m}{E \, \rH}\, ,
\end{equation}
and use the fact that $E$ and $m$ are constants of motion. Then, when the above quantity is smaller than $1$, we have  automatically WCCC preservation (according to Fig.~\ref{fig:Comp1} in the main text). If, instead, there are some particles that satisfy $\hat s/\rH > 1$, they live above the dashed line in Fig.~\ref{fig:Comp1} and potentially can overspin the BH if they reach the overlapping region when $\hat s/\rH > \hat s^{\text{AdS}}_2/\rH$. A necessary condition for this to happen is that also $\hat s/\rH < \hat s^{\text{AdS}}_1/\rH$ must hold, in order to ensure particle absorption by the BH. 

\begin{figure}[t]
    \centering
     \subfigure[]{\includegraphics[width=0.3\textwidth]{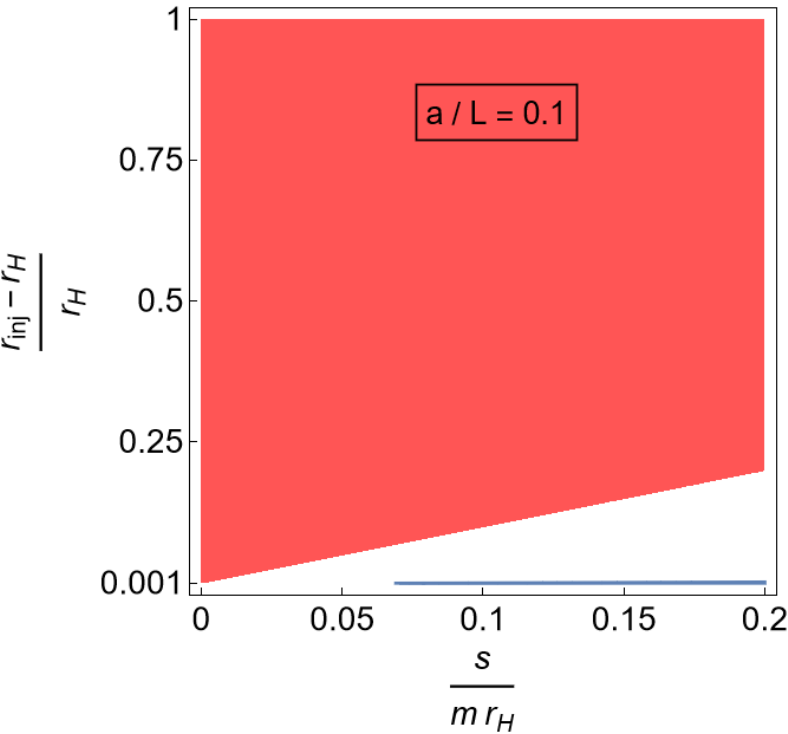}}
    \hfill
    \subfigure[]{\includegraphics[width=0.3\textwidth]{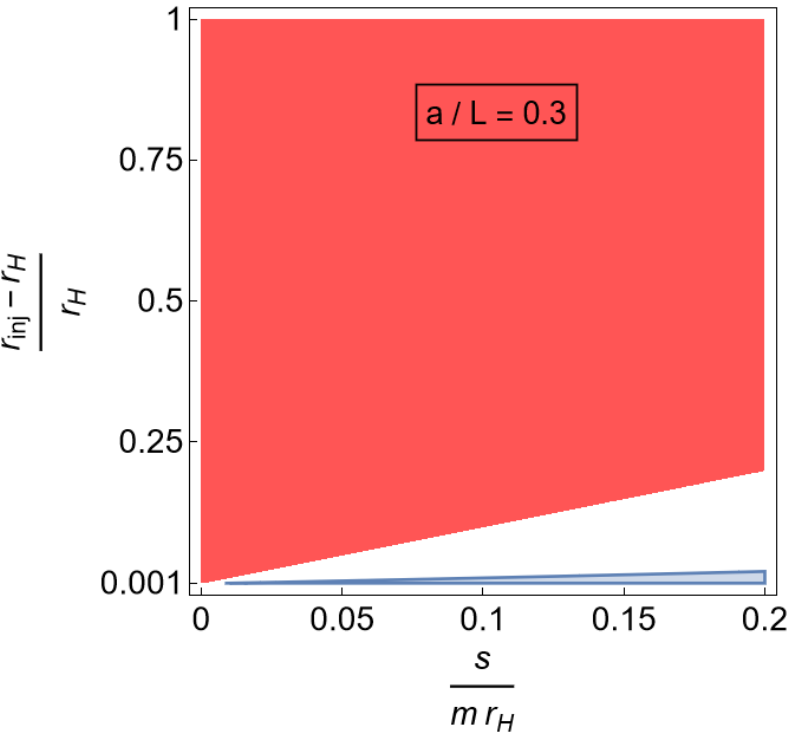}}
    \hfill
    \subfigure[]{\includegraphics[width=0.3\textwidth]{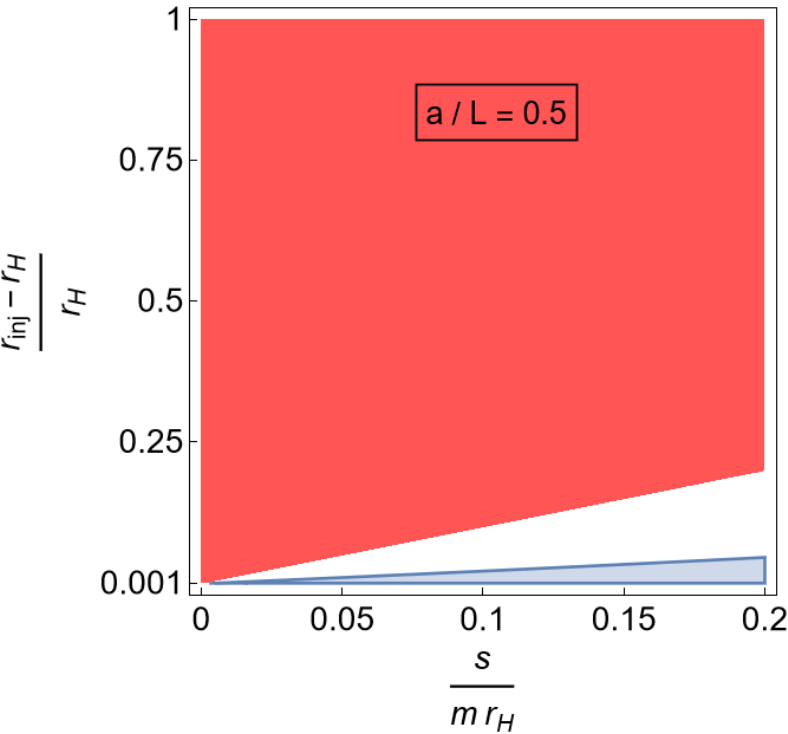}}
    \hfill
    \subfigure[]{\includegraphics[width=0.3\textwidth]{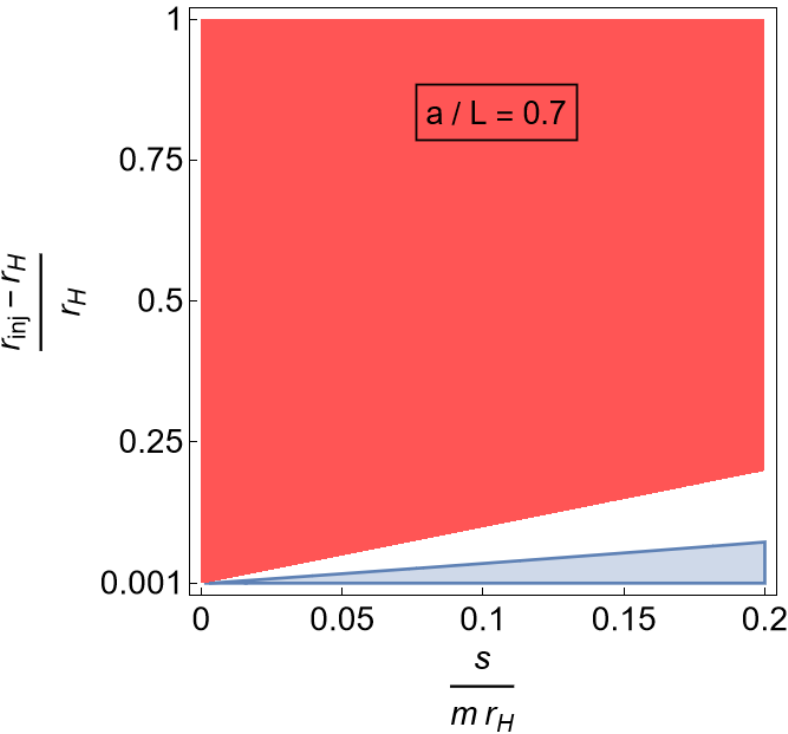}}
    \hfill
    \subfigure[]{\includegraphics[width=0.3\textwidth]{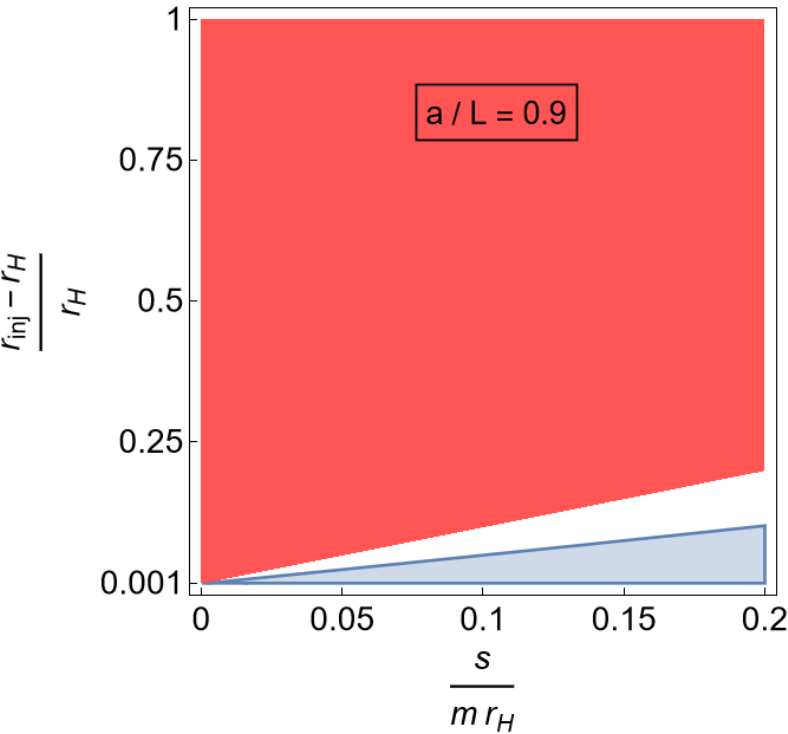}}
    \hfill
    \subfigure[]{\includegraphics[width=0.3\textwidth]{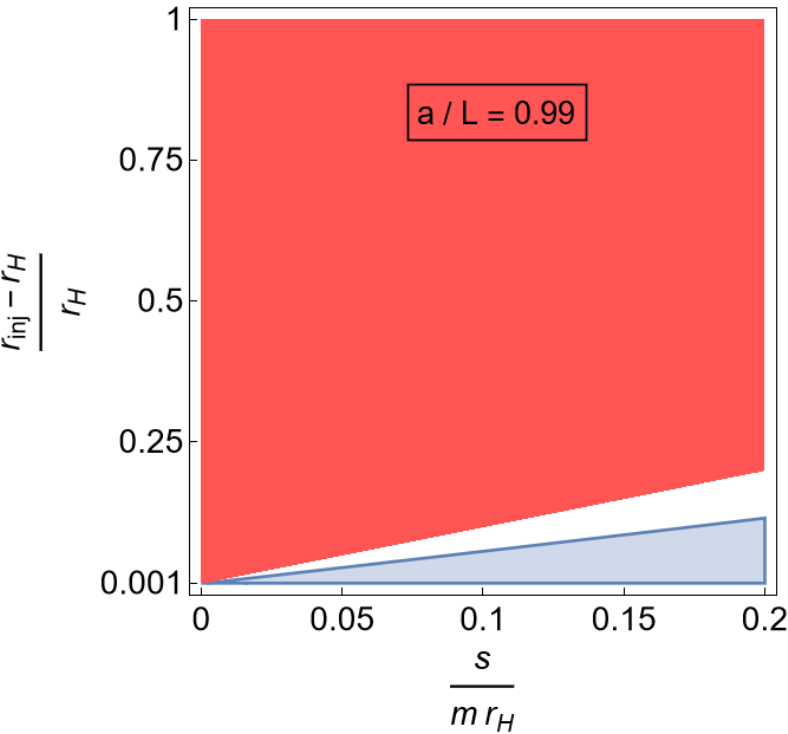}}

    \caption{Blue shaded regions in the parameter space $[s/(m \rH), \, (r_\text{inj}-\rH)/\rH]$ (the normalized spin-to-mass ratio and the normalized initial distance of the particle from the horizon, respectively) represent those where the particle has $\hat s^{\text{AdS}}_1/\rH > \hat s/\rH > \hat s^{\text{AdS}}_2/\rH$ and thus can be absorbed and spin the BH past extremality. Red shaded regions, instead, correspond to regions where the minimum size of the particle (given by $s/m$) is smaller than the distance from the horizon $r_\text{inj} - \rH$. As one can see, there is no overlap between the two regions, meaning that particles able to disrupt the horizon actually have a size bigger than the distance from the horizon itself (they are injected already inside the BH). For each plot, we fixed the value of $a/L$ (determining also the value of $\rH$ through Eq.~\eqref{extremalityconditionKADS} in the main text) and we restricted to particles with $s/m \leq 0.2 \, \rH$, starting (with zero velocity) from a radial distance from the horizon equal to $\ri-\rH \geq 10^{-3} \, \rH$. For $s/m = 0$, one always has $\hat s/\rH < \hat s^{\text{AdS}}_2/\rH$, consistent with no horizon disruption.
    }
 \label{fig:SupplementaryFigure}
\end{figure}

We performed a numerical scan of the parameter space to assess whether such regions, where $\hat s/\rH > \hat s^{\text{AdS}}_2/\rH$, exist. We will consider particles with  $s/m \leq 0.2 \, \rH$ (so that the point-particle approximation, $s/m \ll r_{\text{H}}$, is satisfied). Fig.~\ref{fig:SupplementaryFigure} reports, for different values of $a/L$, the regions in the parameter space where BH horizon disruption occurs. We have checked that $\hat s/\rH < \hat s^{\text{AdS}}_1/\rH$ is always true in the parameter space considered, so all such particles are absorbed.
However, since spinning particles possess a finite size, it is essential to compare this size with their initial distance from the event horizon $r_\text{inj}- \rH$. Specifically, the particle’s proper size must be smaller than this initial plunging distance. In Fig.~\ref{fig:SupplementaryFigure}, we highlight in red the regions of parameter space where the particle’s size ---assumed to be the minimum possible, i.e., $s/m$--- is smaller than its initial distance from the horizon. As evident from the figure, these regions do not overlap with those corresponding to particles capable of overspinning the BH (blue shaded region). This leads to the conclusion that particles seemingly able to violate the extremality bound would, in fact, have to be injected {\it behind} the horizon and are thus not relevant for the WCCC test.
\end{appendix}

\end{document}